# A Portable Brain MRI Scanner for Underserved Settings and Point-Of-Care Imaging

Clarissa Z. Cooley[1,2*], Patrick C. McDaniel[1,3], Jason P. Stockmann[1,2], Sai Abitha Srinivas[1], Stephen Cauley[1,2], Monika Śliwiak[1], Charlotte R. Sappo[1], Christopher F. Vaughn[1], Bastien Guerin[1,2], Matthew S. Rosen[1,2,4], Michael H. Lev[2,5], Lawrence L. Wald[1,2,6*]

[1] Athinoula A. Martinos Center for Biomedical Imaging, Dept. of Radiology, Massachusetts General Hospital, Charlestown, MA, USA.
[2] Harvard Medical School, Boston, MA, USA.
[3] Dept. of Electrical Engineering, Massachusetts Institute of Technology, Cambridge, MA, USA.
[4] Dept. of Physics, Harvard University, Cambridge, MA, USA.
[5] Dept. of Radiology, Massachusetts General Hospital, Boston, MA, USA.
[6] Harvard-MIT Division of Health Sciences and Technology, Cambridge, MA, USA.

*To whom correspondence should be addressed: clarissa@nmr.mgh.harvard.edu, lwald@mgh.harvard.edu

**Abstract**: Access to and availability of MRI scanners is typically limited by their cost, siting and infrastructure requirements. This precludes MRI diagnostics, the reference standard for neurological assessment, in patients who cannot be transported to specialized scanner suites. This includes patients who are critically ill and unstable, and patients located in low-resource settings. The scanner design presented here aims to extend the reach of MRI by substantially reducing these limitations. Our goal is to shift the cost-benefit calculation for MRI toward more frequent and varied use, including improved accessibility worldwide and point of care operation. Here, we describe a portable brain MRI scanner using a compact, lightweight permanent magnet, with a built-in readout field gradient. Our low-field (80 mT) Halbach cylinder design of rare-earth permanent magnets results in a 122 kg magnet with minimal stray-field, requiring neither cryogenics nor external power. The built-in magnetic field gradient reduces reliance on high-power gradient drivers, which not only lowers overall system power and cooling requirements, but also reduces acoustic noise. Imperfections in the encoding fields are mitigated with a generalized iterative image reconstruction technique, that uses prior characterization of the field patterns. Our system was validated using $T_1$, $T_2$ and proton density weighted *in vivo* brain images with a spatial resolution of 2.2 x 1.3 x 6.8 mm$^3$.

## Introduction

Neurological disorders are the 2$^{nd}$ leading cause of death and the leading cause of disability globally [1]. MRI is the reference standard for assessment of these disorders due to its ability to image intracranial anatomy with unparalleled soft tissue contrast. However, large populations of patients are precluded from access to MRI due to its limitations. Most notably, MRI scanners are costly, require special infrastructure, and are immobile. This makes MRI unavailable to patients who cannot be safely transported to the scanner or who are in low-resource settings.

The development of a portable, low-cost MRI device for brain imaging could expand access to MRI neuroimaging and enable point-of-care (POC) diagnostics. In emergency medicine, neuroimaging constitutes the majority of MRI examinations [2]. POC MRI could expedite assessment of neurological emergencies that are not as accurately characterized by CT. For example, POC MRI could detect subtle signs of increased intracranial pressure associated with head trauma, stroke, hematomas, or hydrocephalus. Similar needs exist for critically ill patients in neurological intensive care units (ICUs). These unstable patients can be difficult or unsafe to transport to a fixed MRI scanner that might even be located in a different building [3]. Neonatal imaging introduces related logistical burdens [4] that could be addressed with a POC bedside MRI scanner. Finally, accessible, low-cost MRI could benefit remote low and middle income regions both in the US and abroad, for example to monitor treatment of pediatric hydrocephalus in Sub-Sahara Africa [5]. Overall, a portable MRI head scanner capable of cost-efficient operation outside of a central Radiology department could improve patient outcomes by detecting time-critical pathology and informing immediate clinical management at the point-of-care.

The design paradigm for conventional MRI scanners is fundamentally unsuitable for POC operation. The cost and size of conventional scanners results from their reliance on high-strength, homogeneous, static magnetic fields and switchable linear field gradients [6]. In traditional design, high magnetic fields ($B_0 >= 1.5$ Tesla) are desirable to increase detection sensitivity while high magnetic homogeneity is needed to ensure that the MR image is encoded exclusively by the switchable field gradients. Based on these principles, conventional MRI scanner design has converged on a large superconducting magnet (4-10 tons) requiring high-cost and maintenance-intensive cryogenic components. The switching linear gradient fields are the primary source of acoustic noise (> 130 dB), power utilization (up to 1000 A and 2 kV), and they require water cooling. This combination results in expensive, large, heavy scanners that must be sited in a dedicated suite with special power and cooling services. The complex and potentially dangerous hardware requires highly trained staff to run and maintain the equipment and a safety exclusion zone to prevent projectile accidents with ferrous objects. These aspects contribute to the relative sparsity of MRI scanners compared to other imaging tools - including digital X-ray (DXR), computed tomography (CT), and ultrasound (US) - less expensive systems that can be used in a wider variety of settings. Further, there is a large global disparity in MRI scanner density related to income levels and infrastructure [7].

The need for lower cost and simplified siting of brain scanners has been recognized by the MRI community and has driven recent industrial efforts. This includes the assessment of a 0.5 T whole-body superconducting system [8], the development of a compact superconducting 3 T brain scanner [9], a small footprint, cryogen-free 0.5 T head scanner [10], a 1 T permanent magnet system for siting in the Neonatal ICU [11], a low-field system for dedicated prostate imaging and biopsy guidance [12], and a 64 mT portable brain scanner [13]. In addition to these industrial initiatives, there are academic efforts directed towards more accessible MRI. Low-cost pre-polarized systems have been developed for extremity [14] and brain [15] imaging; the brain imager employed cryogenic SQUID detectors, a 0.1 T pre-polarizing field and ultra-low readout field (0.2 mT). Brain imaging has also been shown in a low-cost, 6.5 mT scanner without pre-polarization and cryogenics, instead focusing on high data-rate image encoding and advanced reconstruction methods [16]. Although these ultra-low field brain scanners are low-cost, they are not portable, and image quality is limited by poor signal-to-noise ratio (SNR) at the ultra-low field strength. A high-field brain scanner has been proposed with a head-only, high-temperature, superconducting magnet [17]. While compact, this design may be inappropriate for truly portable applications given the size of its magnetic footprint and cryogenic requirements. Arrays of permanent magnets have been proposed for low-field portable brain scanners [18–21]. This method is compelling because permanent magnets do not require power or cooling, and the low-field architecture can be configured to have a minimal external magnetic footprint, reducing safety concerns from potential introduced ferrous objects in POC use.

Despite the rapid progress and growing interest in the field, there is no consensus on the best approach for adapting MRI to portable and POC use. To significantly reduce the size, cost, and complexity of the hardware, our design departs from the canonical scanner design (i.e. homogeneous $B_0$ field plus 3 switchable linear gradients). Our unique approach is summarized by 4 points. First, instead of a versatile full body diagnostic device, we focus on a specialized portable design for brain imaging. The small size of the head relative to the torso lends itself naturally to scanner size reduction, facilitating a small diameter, short bore design that fits around the head only. Second, we use a low-field magnet consisting of an optimized array of rare-earth material to generate the static $B_0$ field. The use of permanent magnets capitalizes on the stored magnetic field of these alloys, obviating the need for cryogenics and external current sources. In contrast to the severe SNR penalty at ultra-low field, a low-field magnet in the 50 - 200 mT range provides a workable trade-off between SNR, safety, cost, and footprint required for POC applications. Third, rather than designing the $B_0$ magnet to be homogeneous, we build in a spatial field variation for image encoding. This allows a reduction in magnet size, and it eliminates the need for a traditional "readout" gradient electromagnetic system, reducing the acoustic noise, power, and cooling requirements of the scanner. Finally, we leverage Moore's law by relaxing hardware constraints and addressing the resulting issues with advanced image reconstruction methods, effectively shifting the burden from hardware to software.

Although the sensitivity of the proposed POC scanner is close to that of "low-field" clinical scanners, high-field MRI offers superior image quality and more advanced imaging techniques (e.g. spectroscopy, SWI, DTI). The proposed device, therefore, is not intended to *replace* high-field MRI scanners, but rather, to offer useful MRI diagnostics to populations for whom examination with a conventional, fixed MRI scanner is impractical or impossible, as well as for whom other available imaging modalities, such as ultrasound, provide only limited or suboptimal clinical assessment.

Here, we present the design and validation of the head-only, portable, lightweight, low-field (80 mT) MRI scanner based on a compact permanent magnet array that weighs only 122 kg. Our scanner operates from a standard wall outlet, requires no cooling, and can be mounted on a cart for transportation to the POC. We present the overall scanner and subsystem design, the imaging sequence and reconstruction approach, and *in vivo* brain imaging validation (acquired in an RF shielded room) using $T_1$, $T_2$ and proton density weighted imaging.

**Prototype Scanner**

Figure 1 shows the compact POC scanner located in the RF shielded room with a human subject in position for scanning as well as an exploded view of the in-bore scanner components. From left to right these are the 12-turn single-channel RF transmit/receive coil helmet, the permanent magnet cylinder, the gradient coils, and the RF shield. The total estimated weight of the full scanner system (including magnet, coils, amplifiers, console, and cart) is 230 kg, which can be transported by a single person. If the currently used general purpose prototyping equipment (console, amplifiers, and cart) are replaced with custom efficient lightweight designs, we project a total scanner weight of ~160 kg.

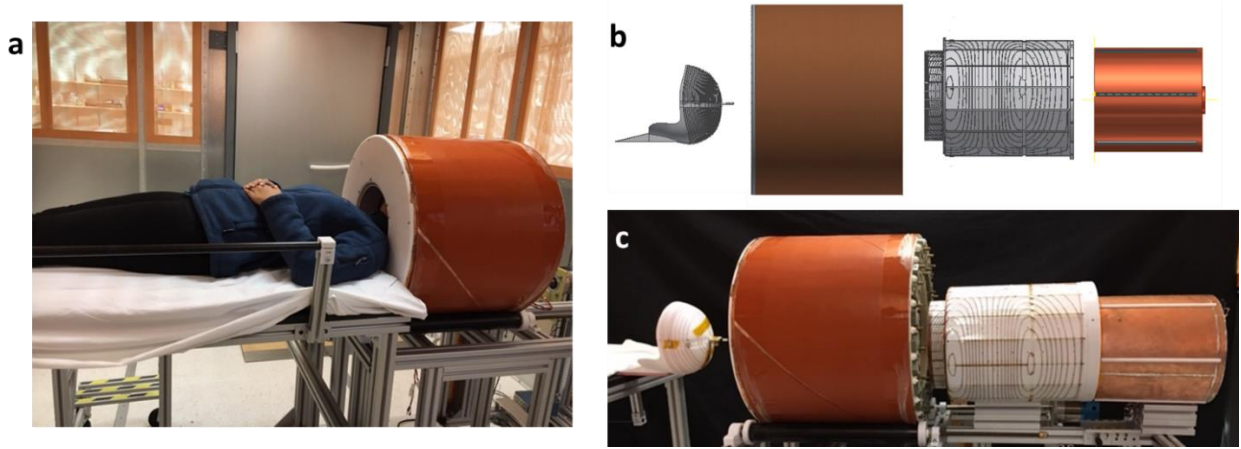

*Figure 1*: Portable MRI brain scanner prototype. (**a**) The scanner main components are inside the 56 cm diameter magnet (orange cylinder). The amplifiers console and computer are not shown. The subject's shoulders remain outside the magnet, allowing for a lightweight, small bore design that fits the head only. The patient table detaches from the scanner cart to facilitate transport. (**b**) Exploded CAD model of the main scanner components (from left to right): spiral transmit/receive RF helmet coil, Halbach magnet cylinder, 2-axis gradient coil, and RF shield. (**c**) Corresponding photo of exploded view.

***Permanent Magnet:*** The head-only permanent magnet consists of a sparse array of NdFeB rare-earth magnets in a Halbach cylinder configuration[18,22]. The ideal dipolar Halbach configuration consists of permanent magnet segments with a magnetization direction that rotates 4π around the cylinder azimuthally[23]. This results in a homogeneous transverse field inside the magnet and zero field outside the magnet. The intrinsically self-shielding design is ideal for portable applications where stray fields pose operational and safety hazards. In addition, unlike other permanent magnet designs, the Halbach magnet does not require a heavy, high permeability (iron/steel) yoke to guide the flux lines, yielding an efficient strength to weight ratio. There is an inherent trade-off between a magnet's size and homogeneity in the imaging region of interest (ROI). For a given volume of permanent magnet material there is also a tradeoff between the diameter of the magnet and the field strength. In order to maintain a small magnet diameter (for portability and field strength), we design the magnet for operation with the subject's shoulders outside the magnet.

In practice, a highly homogeneous Halbach magnet with these geometric constraints is difficult to achieve. Instead of striving to maximize homogeneity in our design, the magnetic field variation is shaped into a built-in field gradient for image encoding. This approach allows a very compact, intrinsically inhomogeneous lightweight magnet and eliminates the need for 1 of the 3 gradient coil systems. Specifically, the built-in gradient replaces the "read-out" gradient system (coil + current driver) which would normally need to overcome the magnet's spurious $B_0$ variation. This would require high power and cooling for conventional encoding approaches within the inhomogeneous magnet. The high-power readout gradient would also produce high acoustic noise during switching. Overall, the built-in gradient paradigm is attractive for a POC scanner as it reduces the magnet cost and size and significantly reduces the full system's power/cooling needs and acoustic noise. However, we note that this scheme reduces flexibility in the choice of pulse sequences.

We allowed an genetic optimization algorithm to perturb the canonical Halbach cylinder design by placing two grades (N42 and N52) of 1" NdFeB cubes to produce a favorable $B_0$ field and built-in gradient in the x direction (see footnote [1]).[18] Figure 2 shows the resulting magnet design with 641 NdFeB 1" cubes arranged in 3 layers of 24 rungs. Figure 2a-b show photos of the superior end of the magnet with the cover removed exposing the ends of the magnet rungs and shim trays (the single row 3rd layer near the shoulders is not visible). The rare-earth cubes are contained within the green, square cross-section, structural fiberglass tubes. Figure 2c,d shows the optimized arrangement of N42 (white) and N52 (grey) magnets and the measured $B_0$ field maps. To improve the gradient field shape, a 2nd optimization stage followed for shim magnet placement with smaller NdFeB elements[26] (visualized in Fig. 2e) using a similar algorithm. Figure 2f shows the improved gradient linearity after shimming. Compared to the orientation of field-maps shown in Figure 2, for imaging experiments, the magnet was rotated 60 degrees to help minimize non-linearities in the FOV.

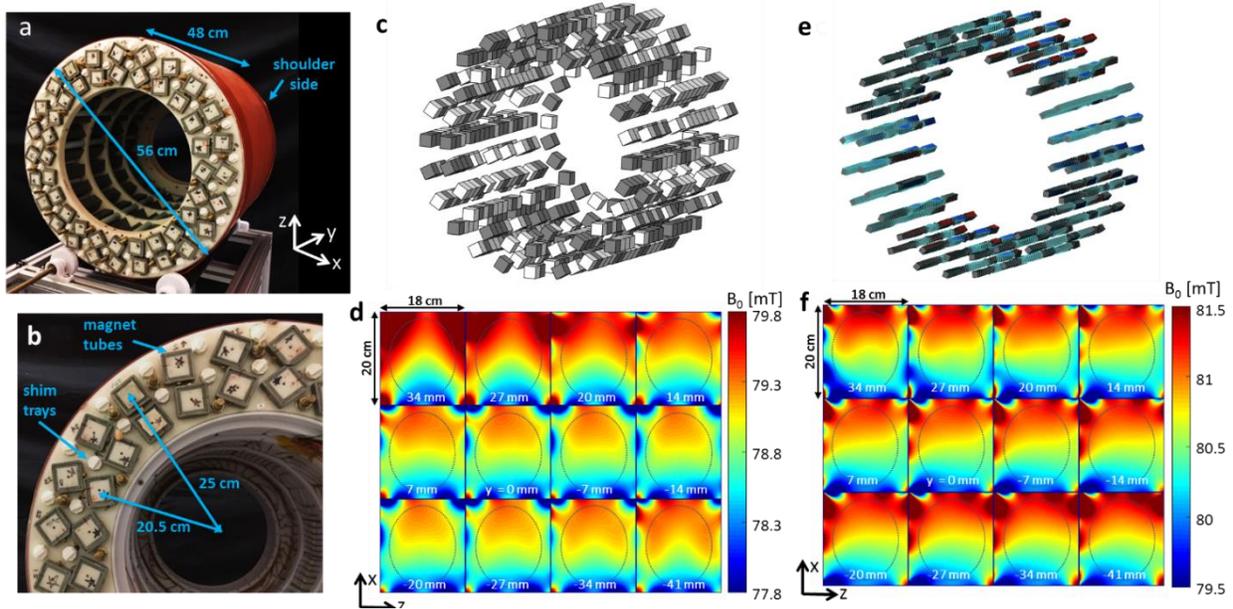

*Figure 2: Permanent low-field magnet design. (a) The $B_0$ = 80 mT cylindrical Halbach magnet has an outer diameter = 56 cm, length = 48 cm, total weight = 122 kg (80 kg of rare earth material). Photo shows the superior side ("service end") of the magnet with a 35.3 cm diameter opening. The inferior side of the magnet (shoulder side) has a 27 cm bore opening due to the 32 cm dia. ring of 1" "booster" magnets near the shoulders placed to alleviate the field fall-off. (b) Close-up photo of superior end of magnet. The 1" NdFeB magnets are contained within the square cross-section fiberglass tubes. The two main magnet layers are at radii 20.5 cm and 25 cm. The white plastic shim trays contain the addition of smaller NdFeB magnets to further optimize the magnet field. (c) CAD model showing the distribution of N52 grade (grey) and N42 grade (white) NdFeB 1" cubes comprising the Halbach magnet optimized for a built-in monotonic read-out encoding field in the x direction. (c) Measured field-map in the axial 18 x 20 cm planes for the constructed magnet distribution prior to shimming. The 17 x 14 cm ovals outline approximate brain dimensions. (e) CAD model of shim magnet distribution for fine-tuning of the field. The smaller shim magnets axial position was fixed, but size and the dipole direction were varied. (f) Measured field-map of the shimmed magnet, showing an improvement in the field linearity in x.*

The constructed magnet assembly has length = 49cm, outer diameter = 57cm, inner diameter = 35cm, and bore access diameter at the shoulders of 27cm. The magnet used 80 kg of rare earth material and the constructed assembly weight was 122 kg. The $B_0$ field averaged 80 mT over the 20 cm DSV target volume and contained a built-in readout gradient of 7.6 mT/m. On average the pull-force on a ferrous object equaled its weight at ~13 cm from the bore opening and ~1 cm from the outer cylindrical surface, demonstrating a considerably smaller safety footprint than conventional high-field MRI magnets.

---

[1] Compared to previous publications [18,22,24–26], the coordinate system was changed to adhere to the more traditional use of "x" for the readout gradient direction.

*Gradient coils:* While the magnet's built-in field variation is used for image encoding in the x dimension, we used the switchable gradient coils shown in Fig. 3 for phase encoding in the y and z direction. Previously, we introduced alternative encoding methods that further reduced the need for switchable gradient systems, specifically the combination of generalized projection imaging by rotating the Halbach magnet[22,25] and either a phase-encode gradient coil or RF encoding method such as Transmit Array Spatial Encoding (TRASE)[24] for encoding along the axis of the cylinder. While these methods can further reduce or eliminate the need for gradient power amplifiers (GPAs), they also require additional hardware. Moreover, the use of switchable gradients for phase-encoding within a spin-echo sequence proved more robust to image artifacts. Although we employ switchable gradients to encode in two directions (y and z), minimal power is needed compared to conventional scanners. The power reduction arises from two sources. First, unlike the readout gradient, phase encoding gradients need not dominate the $B_0$ inhomogeneity ($\Delta B_0$) in a spin-echo sequence since the $\Delta B_0$ phase dispersion is refocused in the spin-echo. Second, the permanent magnet design supports the use of efficient gradient coils without a shielding layer (secondary gradient coil winding). In the imaging experiments described here, peak currents of 9A and 4.5A were used to the drive the z and y gradient respectively.

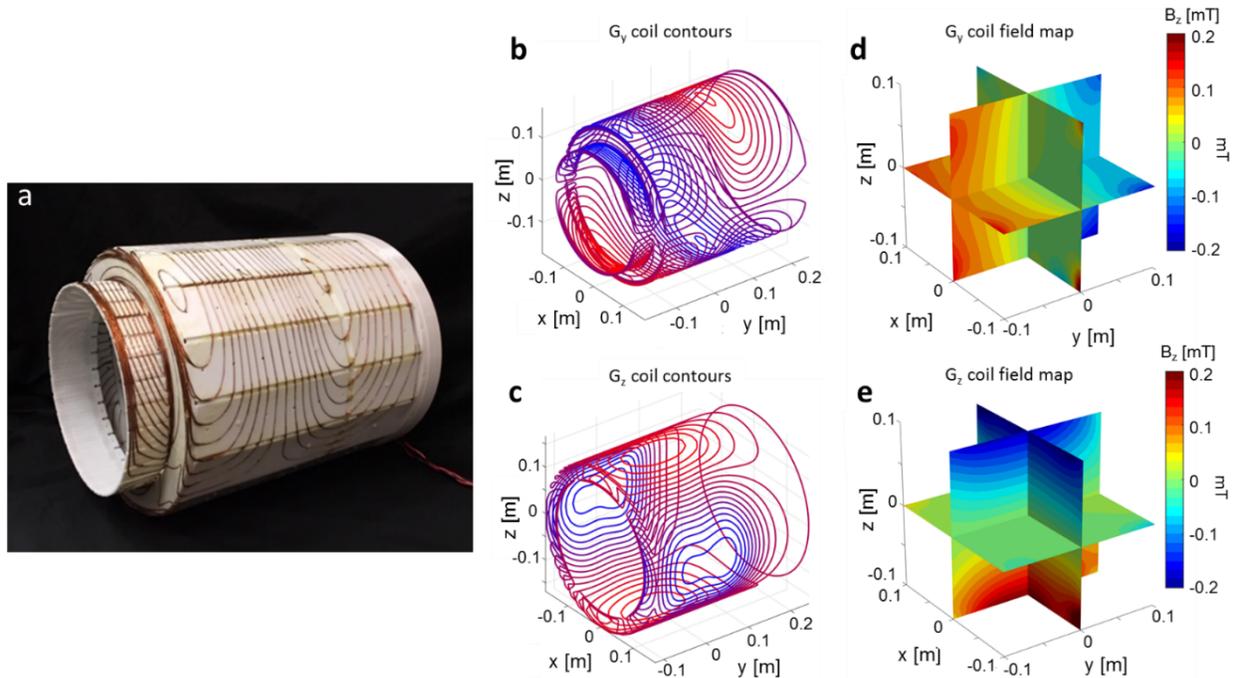

*Figure 3:* Gradient coil design. (*a*) $G_y$ and $G_z$ gradient coils with wires press-fit into a tiered cylinder 3D printed former. The $G_y$ gradient coil is on the outer surface and $G_z$ is on the inner surface. The tiered shape allows for maximum diameter (34.8 cm) and length (42.7 cm) within the magnet. (*b,c*) Gradient coils' current density contours designed with a BEM stream function method optimized for linearity in the 20 cm ROI. (*d,e*) The measured gradient coil field maps for 1 A of drive current in the coils. The $G_y$ and $G_z$ coil efficiencies were 0.6 mT/m/A and 0.8 mT/m/A respectively.

*Sequences:* The low $B_0$ field and built-in gradient poses unique MRI spin manipulation problems and sequence considerations. Because the RF frequencies of the transmit ($B_1$+) and receive ($B_1$-) magnetic fields are proportional to the inhomogeneous $B_0$ field, the 20 cm ROI encapsulates a Larmor range of 3.35-3.43 MHz (80 KHz bandwidth). Traditional "hard" $B_1$+ pulses would require high RF power levels to manipulate all the spins in the wide Larmor frequency bandwidth. Instead, we employ frequency-swept chirped $B_1^+$ pulses for excitation and refocusing, which cover a large bandwidth and are less susceptible to $B_1^+$ amplitude variation [24,27].

The built-in gradient precludes the standard formation of a gradient echo and thus limits the MRI acquisition method to spin-echo based sequences. Compared to high field MRI, the lower $B_0$ field leads to low RF heating and longer spin coherence times (T2 relaxation times). We take advantage of these properties to enable an efficient acquisition sequence using long multi-echo Rapid Acquisition and Relaxation Enhancement (RARE) [28] volumetric spin-echo sequences (Fig. 4).

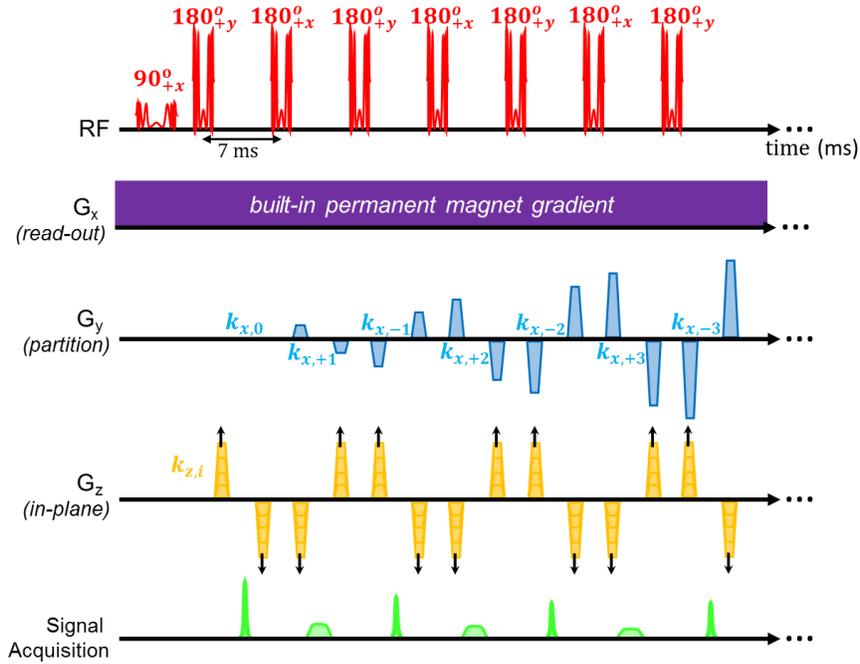

*Figure 4:* MRI pulse sequence diagram. The 3D RARE (Rapid Imaging with Refocused Echoes) pulse sequence is shown for proton density (PD) weighted sequence. The RF applies the 90 degree excitation chirped pulse (3.2 ms, 100 kHz sweep) followed by a train of 180 degree chirped refocusing pulses (1.6 ms, 100 kHz sweep). The phase of the pulses follows a phase cycling scheme that prevents mixing of the resulting FID and Spectral echoes. The $G_x$ readout gradient is the built-in permanent magnet encoding field, and therefore is continuously applied throughout the acquisition. The $G_y$ gradient produces phase encoding blips that vary along the echo train for partitioning data in the y dimension completing the 23 encodes in each shot. The $G_z$ phase encoding blips are incremented shot-to-shot requiring 97 TR periods to complete the encoding. The Signal Acquisition alternates between the narrow "FID echoes" and wider "spectral echoes". The sequence is converted to $T_1$-weighting with the addition of an initial inversion pulse. In the $T_2$-weighted sequence, the ordering of the $G_y$ phase encoding blips are re-arranged so that the center of k-space is captured at $TE_{eff}$ = 167 ms.

**Image Reconstruction**

Traditional MRI image reconstruction relies on the use of linear encoding fields to reconstruct k-space data using the Fast Fourier Transform (FFT). Although we optimized the gradient encoding fields for linearity, the compact nature of the system limits linearity in the ROI, particularly towards the periphery of the permanent magnet gradient – $G_x$. Non-linear encoding fields can lead to image aliasing and "encoding holes" [22,29,30], which can sometimes be alleviated with the use of multi-coil receive arrays. However, if the encoding fields are monotonic, the non-linearities will translate to more benign geometric distortion and variable resolution in the image. If relatively small, this distortion can be corrected using a model-based generalized image reconstruction technique[31] utilizing *a priori* knowledge of the fields. These generalized techniques employ a forward model of the time domain signal evolution in response to the known encoding non-linear fields[29,32,33,30,22]. Our encoding model uses the measured field-maps of the built-in readout gradient (Fig. 2f) and the gradient coils (Fig. 3c,e), and models the time domain encoding process of our 3D RARE sequences.

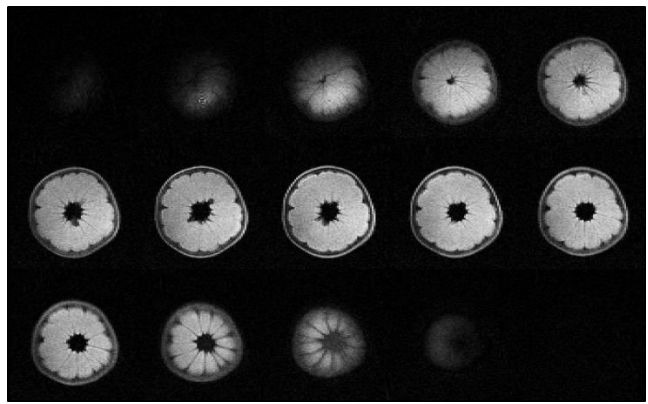

*Figure 5:* 3D image of a 11 cm diameter grapefruit is shown using a T2-weighted 3D RARE sequence. Acquisition parameters: 2500ms/167ms, acquisition time = 16:10 min (4 averages). Image resolution = 2.2 x 1.3 x 6.8 mm$^3$.

We solve for the image using an iterative conjugate gradient algorithm implemented in MATLAB (Mathworks, Natick, MA) with GPUs. Figure 5 shows a T2-weighted 3D image of a ~11 cm diameter grapefruit with no visible distortion. Figure 6 shows brain images of 3 healthy subjects with an image resolution of approximately 2.2 x 1.3 x 6.8 mm. This approximate resolution is calculated from a linear fit of the field-maps but resolution actually varies slightly over the FOV due to the encoding field non-linearities. The proof-of-principle images were acquired in an RF shielded room. The top 3 rows show PD, $T_1$, and $T_2$-weighted contrasts in the same subject (S1), followed by $T_2$-weighted images in two additional subjects (S2 and S3). The S3 FFT image (bottom row) was formed using a conventional reconstruction technique on the S3 $T_2$ data, which assumed linear field gradients instead of the measured non-linear fields. Comparing this to the generalized reconstruction technique demonstrates the distortion improvements achievable using *a priori* information of the encoding fields. However, some image distortion remains towards the periphery where $G_x$ is less linear. With the exception of S3 T2 and S3 FFT, each image was acquired in ~ 10 minutes. S3 T2 (S3 FFT) was acquired in 19 minutes to allow for more averages and higher SNR.

**Discussion**

Our portable scanner is capable of generating standard brain MRI contrasts found on low-field clinical scanners – including T1, T2, inversion recovery (IR)-prepped T2, proton density (PD), and diffusion weighted images (DWI) – that are routinely used for detection, diagnosis, and monitoring of clinically important brain pathology. The scanner offers superior soft-tissue contrast resolution compared to other imaging modalities available for POC use- such as ultrasound (US), digital X-ray (DXR), and computed tomography (CT) – which are additionally limited by acoustic shadowing (US), beam hardening artifacts (DXR, CT) from bone and calcified structures, ionizing radiation (DXR, CT), and poor ability to distinguish certain central nervous system anatomic structures (e.g., gray versus white matter, subdural versus extradural spaces). Although both the spatial resolution and sensitivity of this scanner are less than that of a high-field MRI, the performance is sufficient to detect and characterize serious intracranial processes at the POC, such as hemorrhage, hydrocephalous, infarction, & mass lesions. Indeed, our portable, compact, affordable device could extend the reach of MRI to answer critical, time-sensitive questions in settings where MRI is not currently available, including urgent care centers, emergency rooms, intensive care units, sports arenas, oncology clinics, remote field hospitals (e.g., for military & humanitarian assistance missions), and perhaps even ambulances.

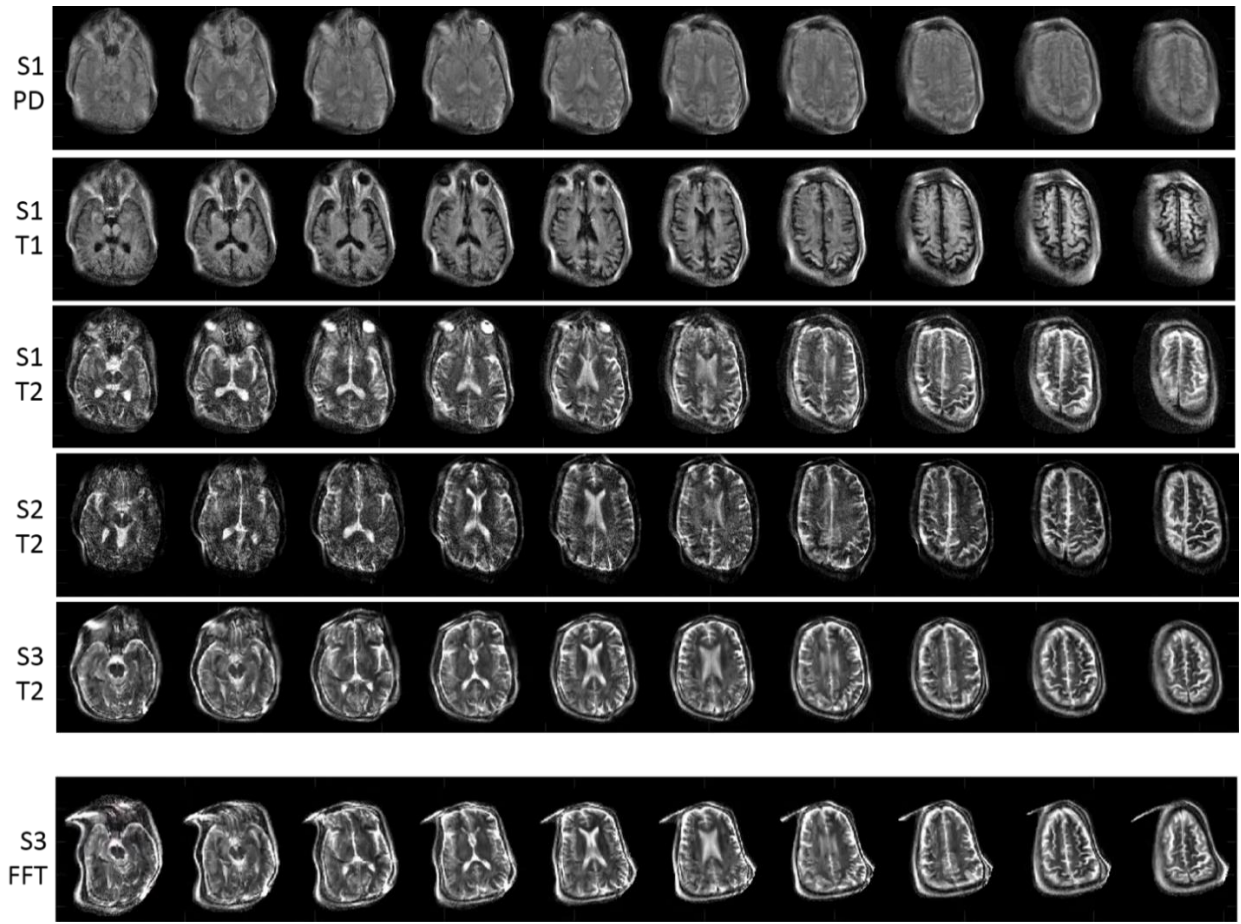

*Figure 6*: 3D T2, T1 and PD-weighted images of the brain in healthy adult volunteers. A subset of the acquired 23 partitions are shown. Image resolution ~ 2.2 x 1.3 x 6.8 mm$^3$. The first 5 rows show images reconstructed with the generalized forward-model based reconstruction method. S1 PD: (subject 1, male 63 years old) PD images acquired with 3D RARE, TR/TEeff = 2900ms / 14ms, acquisition time = 9:24 min (2 averages). S1 T1: (subject 1) $T_1$-weighted images acquired with inversion prepped 3D RARE, TI/TR/TEeff = 400ms / 1830ms / 14ms, acquisition time = 11:46 min (4 averages). S1 T2: (subject 1) $T_2$-weighted images acquired with 3D RARE sequence, TR/TEeff =3000ms/167ms, acquisition time = 9:42 min (2 averages). S2 T2: (subject 2, male 63 years old) $T_2$-weighted images acquired with 3D RARE sequence, TR/TEeff =3000ms/167ms, acquisition time = 9:42 min (2 averages). S3 T2: (subject 3, female, 53 years old) $T_2$-weighted images acquired with 3D RARE sequence, TR/TEeff =3000ms/167ms, acquisition time = 19:24 min (4 averages). S3 FFT: the S3 T2 data reconstructed with a conventional FFT reconstruction instead of the generalized method. This last image demonstrates the geometric distortion that results from the non-linear encoding fields when the field-maps are not included in the reconstruction model. The measured SNR in the images were SNR = 127, 80, 68, 65, 124 for the image acquisitions in rows 1-5 respectively.

Although our proposed MRI scanner design fulfills many of the potential requirements for a POC brain imaging device, several considerations require additional attention. Currently, for example, image distortion is not fully addressed by our reconstruction algorithm. Specifically, signal aliasing may be occurring due to the curvature of the field-map isocontours, rendering the encoding fields non-orthogonal - a situation not included in the model. Remaining geometric distortions may also have contributions from measurement errors in the encoding field maps. These distortions become more marked further from isocenter, where the non-linearities are more severe. The encoding field non-linearities and their effect on the image are analyzed in Figure 7. The measured $B_0$ field ($G_x$ gradient) in a central axial partition, for example, gives rise to the higher error levels near the periphery of the ROI. Figure 7 also shows the resulting spatial deformations expected in the image formed from a simple the FFT reconstruction; these same maps are also shown for the $G_z$ and $G_y$ gradient coil fields. Here the errors and spatial deformation are less severe, motivating the potential use of the FFT in those dimensions to decrease the computational burden of the generalized reconstruction.

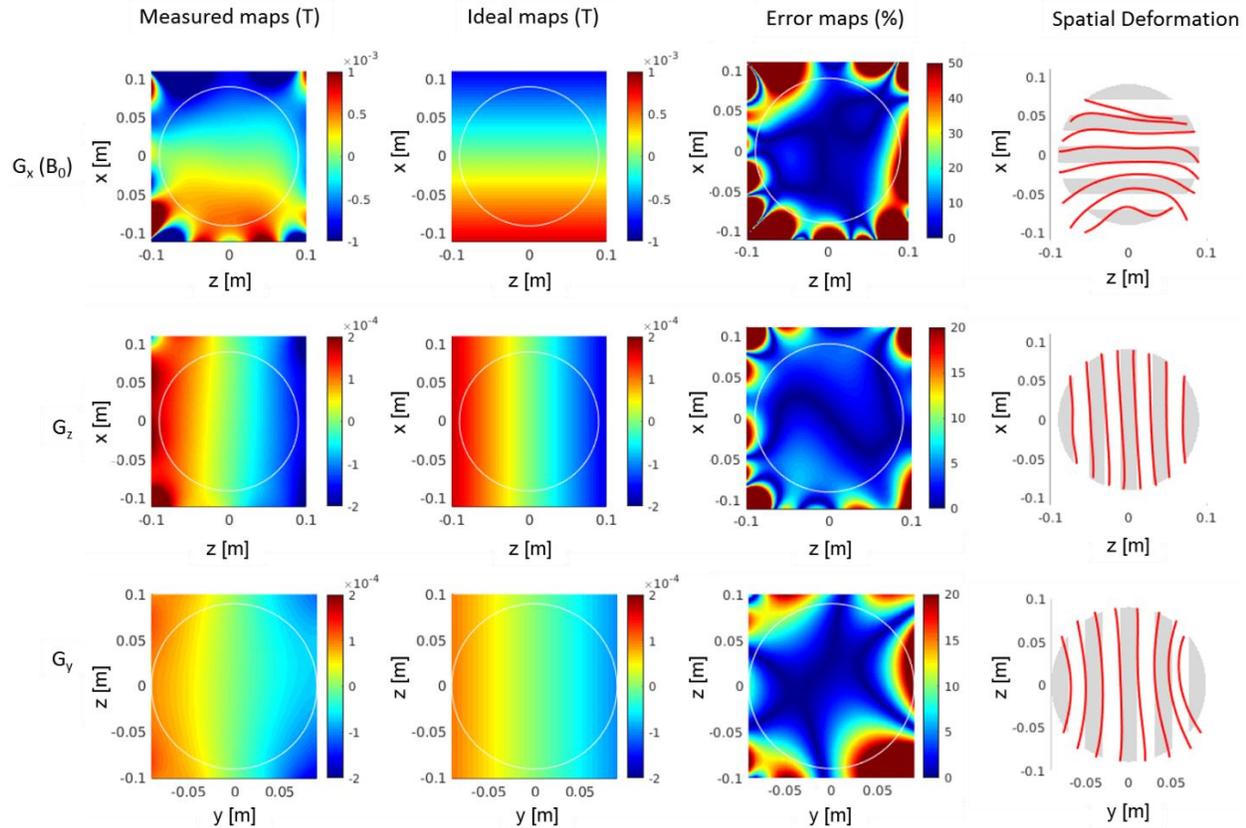

*Figure 7:* *Analysis of the measured encoding field-maps ($G_x$, $G_z$, and $G_y$) in the central field-map slices. The ideal maps are calculated as a linear fit to the measured maps. The error maps show the percent difference between the measured maps and ideal maps. The color range is higher (up to 50%) for the $G_x$ gradient. Spatial deformation maps show the resulting image distortion that occurs when the ideal linear map is assumed (instead of the measured map). The non-linearities and spatial deformation are most severe in the $G_x$ encoding map, which is generated by the built-in permanent magnet gradient. The $G_x$ analysis shows high errors the periphery and severe spatial deformation approaching signal singularities in some locations. In contrast, the gradient coil maps ($G_z$ and $G_y$) errors and spatial deformation maps are more benign.*

The high temperature coefficient of magnetic remanence (~-0.1%/C) and coercivity (~-0.5%/C) in NdFeB material[34] results in variation in the $B_0$ field with room temperature, and could contribute to errors in the reconstruction model. Before each data set is acquired, the center Larmor frequency is set, reducing large off-resonance effects from temperature drift. Further, a global $B_0$ offset variable is adjusted in the reconstruction model to account for differences in the experimental $B_0$ field compared to the previously measured $B_0$ field maps. However, we currently assume that there are no significant temperature changes on the time scale of each image acquisition (~10 minutes). To improve the accuracy of the reconstruction model, field probes can be used to track global [22] or local changes in the $B_0$ field during the data acquisition.

Future iterations of the scanner design could focus more on industrial patient interfacing and workflow considerations. The compact size of the scanner results in a tight fit around the head, requiring special attention to the mechanical design of this area, including entrance and exit patient positioning

(especially for intubated and highly monitored patients), and airflow and monitoring within the bore. The volunteers imaged with the scanner, however, found it to be comfortable during their >45-minute acquisition session. Our head-only magnet design, moreover, improves patient comfort by eliminating confinement around the body. Further, our design allowed acoustically quiet operation, eliminating the need for ear plugs.

The preliminary images presented here were acquired in an electromagnetically shielded room, like that of traditional scanners. This RF shielded scanner suite eliminates electromagnetic interference (EMI), which can otherwise degrade image quality (signal to noise ratio, SNR) and can introduce artifacts. In our portable system, although a copper shield was placed between the gradient coils and RF head coil, this was insufficient to fully eliminate EMI in human imaging. The openings in these shields are small compared to the relevant wavelength of the system's MRI signal (89 m), and the built-in shield was sufficient to prevent EMI when imaging small phantoms or fruit (Fig. 5). When imaging human subjects, however, the body parts outside of the shield act as an antenna, which conducts EMI into the MRI receiver coil. For our pilot human imaging validation, therefore, we operated the scanner inside a traditional Faraday cage to eliminate this source of image degradation. Work is ongoing, however, to actively record and remove interference using external pick-up coils that can monitor environmental EMI during imaging [35,36].

Ongoing efforts also include optimizing RF coil designs to increase SNR and extend FOV to more inferior brain regions, as well as testing and validation with specific brain pathologies, such as small vessel white matter disease, in addition to testing healthy subjects. Preliminary work also suggests that diffusion weighted imaging – critical to certain applications, such as acute stroke detection - is also possible with our unconventional scanner architecture[37].

**Outlook**

Our scanner design could serve as a foundation to develop and clinically validate portable MRI devices for affordable, point-of-care detection, assessment, and monitoring of diverse medical applications. In addition to the diagnostic whole-brain imaging applications discussed, for example, our architecture could be minimally modified for extremity and neonatal imaging. Moreover, extending the general concept of liberating MRI design from traditional constraints might lead to even more exotic designs with extreme portability, such as hand-held devices. Devices that generate limited FOV images or profiles directly under a single-sided scanner[12,38] could be used for diverse "real-time" emergency and urgent care indications – such as detection, delineation, & serial monitoring of soft tissue pathologies (e.g., pleural effusions, extremity abscesses requiring drainage, or subdural/epidural hematomas), or to provide guidance for interventional procedures (e.g., catheter placement, lumbar punctures[35], or biopsies[12]). Such devices have the potential to complement or replace the roles of other (often suboptimal or more limited) portable imaging modalities.

In conclusion, we introduced a new MRI scanner architecture based on a compact, lightweight, low-field permanent magnet array, with built-in field variation for MRI readout encoding and efficient electromagnetic gradient coils for phase encoding. Our design leverages advanced image reconstruction methods to correct for magnetic field imperfections, freeing the hardware from traditional constraints. Unlike conventional MRI scanner designs, this approach could allow for POC operation due to the magnet's modest size, lack of cryogenics, and the intrinsic safety of the low-field, magnetically self-shielded Halbach configuration. Both mobility and POC potential are also facilitated by the low power consumption and low acoustic noise afforded by our built-in readout gradient. The presented *in vivo* brain images demonstrate the potential of the scanner for clinical application at the POC, which could expand access of MRI to patient populations now underserved by traditional MRI limitations.

## Methods

***Permanent magnet construction :*** The head-sized permanent magnet was designed using a genetic optimization framework previously described by Cooley et al [18]. The dimensions and basic geometry of the sparse Halbach magnet were determined based on human anatomy and tradeoffs between field strength and size. The region of interest is defined as a 20 cm sphere with the isocenter at 17.8 cm from the inferior end of the magnet (constrained by the shoulders). The magnet is asymmetric, extending 27.9 cm above isocenter (superior direction) to improve homogeneity. A "booster" ring of magnets is added near the shoulders to compensate for the field "fall-off" effects here.

The Halbach cylinder is made up of square cross-section permanent magnet rungs divided into 2 full layers at diameters = 41 cm and 50 cm. Each layer contains 24 rungs that are 45.7 cm in length. The additional Halbach "booster ring" near the patient's shoulders has a diameter = 32 cm and length = 2.54 cm (1 magnet row). NdFeB rare earth magnetic material was chosen because of its high remnant flux density, coercivity and lower cost compared to SmCo. The magnet was constructed with stock 1" NdFeB cubes (Applied Magnets, Plano, TX). The use of standardized 1" NdFeB cubes eases the cost and construction of the magnet and the sparsity of the design reduces the cost and weight (albeit at the cost of field strength).

In the full magnet geometry, there are 888 predetermined potential locations for the NdFeB cubes. A Genetic Algorithm determined the placement of either N42 and N52 grade NdFeB cubes or plastic spacers at each location. The optimization was constrained to produce a mean $B_0$ >70mT with a monotonic encoding field and reasonable total field range[18]. The resulting design, shown in Figure 2, contains 342 and 299 N42 and N52 NdFeB cubes respectively (~80 kg of NdFeB material).

The non-magnetic housing for the NdFeB material uses 1" cross-section structural fiberglass square tubes that contain the magnet material. The rungs are mechanically supported by seven 1.27cm-thick ABS plastic rings with waterjet cut square holes rotated in the Halbach configuration. The design also contained 48 octagonal holes meant to hold trays of smaller shim magnets. Threaded brass fastening rods and fiberglass dowel spacers increased structural integrity. After full assembly of the mechanical housing, a pushing jig was used to populate the NdFeB cubes into the corresponding rungs. The cube magnets repel each other within the rungs, so the jig was needed to temporarily extend the fiberglass rung length so the magnets could float apart. Then the jig was used to push all the magnets into contact within the housing and bolt an end cap on the tube. This was repeated for all 48 rungs. The NdFeB material was handled with caution as serious injury could result from the forces between the NdFeB cubes. The resulting magnet structure has a length = 49cm, outer diameter = 57cm, inner diameter = 35cm, bore access diameter = 27cm and weight = 122 kg.

A field-mapping robot was constructed to measure the field pattern in the permanent magnet and gradient coils. The robot was based on a modified "build-your-own" CNC router kit (Avid CNC, North Bend, Washington, USA) which rastered a 3-axis gaussmeter probe (Metrolab Technology, Geneva, Switzerland). MATLAB software was used to simultaneously control the stepper motors to traverse the probe through the magnet's ROI and record the gaussmeter field measurements.

The field at construction was dominated by a 1st order field variation, but the existing non-linearities caused severe image distortion and some singularities (aliasing). Therefore, a target-field shimming iteration was used to refine the built-in encoding field of the magnet[26]. This was achieved with an optimized population of the 48 shim trays (each containing 42 shim magnet locations (for up to 6.35mm NdFeB cubes). The orientation (dipole direction) and size of NdFeB shim magnets at each of the 2016 potential shim locations was optimized to minimize the RMSE deviation from an ideal linear "target" gradient in the ROI. This calculation used an interior-point MATLAB optimization with each shim magnet modeled as an ideal magnetic dipole. The varying size of the resulting dipoles was practically realized by gluing smaller magnet pieces together. The resulting shape and orientation of each shim magnet was designed into the 3D printed shim trays.

***Gradient coil construction:*** The gradient coils were designed to create linear target field gradients in the y and z direction in the imaging ROI[26,39]. The mechanical surfaces of the 2 coils were predetermined to be on the inner and outer surface of a tiered cylinder former designed to fit snuggly inside the magnet (length = 42.7 cm, diameter 1 = 34.8 cm, diameter 2 = 26.4 cm). The current stream functions of the coils were optimized on a 20 cm diameter ROI using a stream function BEM solver based on a published toolbox[40]. The achievable current density at the truncated end of the coil (shoulder side) is limited by practical densely of the windings in this area which proved to be the main constraint limiting the coil's efficiency and linearity. Based on the optimized current stream function, the coil winding patterns were designed for a target gradient efficiency of 0.7 mT/m/A and a resistance of < 2 ohms. AWG 18 wire was press-fit into wire winding pattern grooves in a 3D printed former. The field-mapping robot was used to measure the field pattern when each coil was driven with 1 A inside the magnet. The resulting gradient efficiencies and inductances/resistances were determined to be 0.575 mT/m/A and 0.815 mT/m/A and 514 uH/1.9 Ohms and 336 uH/1.2 Ohms for $G_y$ and $G_z$, respectively. In the imaging sequences described here, less than 10 A peak current is used at a low-duty cycle (3-5%) allowing for passive air-cooling. The low power requirements of the gradient systems will allow for the future integration of very low-cost, low-power, small-footprint op-amp based drivers[41].

*RF coil construction:* The RF coil (Fig.1b,c) is based on a compact spiral helmet design[42] used for transmit and receive with a passive transmit-receive switch. The coil is wound on a tightly fitting helmet former of inner dimensions: 21 cm (anterior-posterior), 17 cm (medial – lateral)[42]. The windings extend 10.7 cm from the top of the head. The close-fitting spiral pattern provides favorable RF receive efficiency and sensitivity. However, when the windings are uniformly distributed on the helmet, the resulting $B_1$ field is inhomogeneous with a 87% higher field produced at the top of the head compared to the bottom in simulation. When used as a transmit-receive coil, the inhomogeneous nature of the resulting $B_1+$ pattern causes variable flip angles in the brain and image artifacts. To improve the $B_1+$ homogeneity, the winding distribution was empirically adjusted using Biot-Savart simulations, resulting in a total of 12 asymmetric windings with a higher turn-density near the bottom of the coil. This reduced the $B_1$ range in the helmet by 79% compared to the uniform winding design. The coil was constructed on a 3D printed polycarbonate former with winding grooves. The non-uniform turn distribution was wound with Litz wire (AWG 20 5/39/42, New England Wire, Lisbon, NH) and tuned and matched to 50 ohms at the system's 3.39 MHz Larmor frequency with non-magnetic capacitors.

*Other hardware:* A passive crossed diode-based, lumped element quarter-wave 50 ohm transmit-receive switch is used with the RF coil. Reception used a low-noise 50-ohm input impedance, 37 dB gain pre-amplifier (MITEQ model AU-1583, Hauppauge, NY, USA) and a 24 dB second stage amplifier (Minicircuits model ZFL-500LN+, Brooklyn, NY, USA). Additional hardware includes: a Tecmag Bluestone MR console (Houston, TX, USA), AE Techron 7224 gradient amplifiers (Elkhart, IN, USA), a 2kW peak-power RF power amplifier (Tomco Technologies model BT02000-AlphaS-3MHz, Stepney, Australia), and patient table constructed from T-slot aluminum extrusions. While this equipment is well-suited for prototyping and validating the scanner design, the console, gradient amplifiers, RF amplifier, and patient table could be replaced with custom designs that prioritize cost and weight[41,43,44].

*Acquisition method:* The permanent magnet readout encoding field is always on, causing an inhomogeneous $B_0$ field ($\Delta B_0/B_0 \sim 2\%$) and a wide Larmor frequency bandwidth in the ROI (~80 kHz). For wide bandwidth RF excitation and refocusing in the spin-echo train, shaped frequency-swept RF pulses (WURST pulses) were transmitted instead of rectangular single-frequency pulses (hard pulses). This use of WURST pulses for MRI in an inhomogeneous field has been previously described [24,27]. Although similar to a rectangular chirped pulse, WURST pulses have a soft taper on the rising and falling edge of the pulse to reduce ringing artifacts and achieve a smooth transition at the edges of the frequency band of excited spins. The excitation and refocusing pulses used in our sequences are 3.2 and 1.6 ms long (respectively) with a 100 KHz linear frequency sweep and a WURST-40 amplitude envelope. The simulations in Figure 8 demonstrate the bandwidth coverage of the pulses and the robustness to $B_1$ variation in the refocusing pulses.

   The linear frequency sweep of the RF pulses imparts an undesired quadratic phase on the spins across frequency. When the background field gradient is held constant during excitation and refocusing, the quadratic phase can be removed from odd-numbered echoes by setting the frequency sweep rate of the refocusing pulse to be twice as fast as that of the excitation pulse[45]. The resulting quadratic phase cancellation in the odd echoes of the RARE echo train results in "FID echoes" (classic spin-echoes). However, the even-numbered echoes contain the quadratic phase, resulting in "spectral echoes" in which different spin isochromats refocus at different time points. Confounding mixing of the FID and spectral echoes can result from flip angle errors. To alleviate this, phase cycling of the RF pulses (alternating 0 and 90 degrees) is used to form two spin coherence pathways[27]. Although there are schemes to combine data from the two types of echos[24,27], we reconstruct only the spectral echoes to limit data inconsistency.

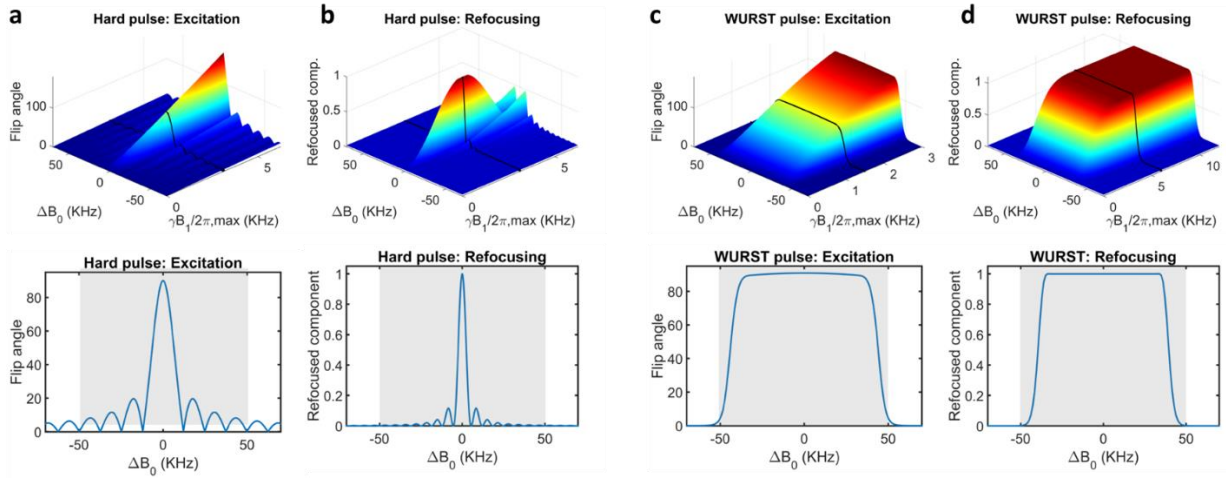

*Figure 8:* RF pulse simulations. The excitation and refocusing pulses are shown as a function of $B_1+$ and $B_0$ off-resonance (3D plots) as well as plots of off-resonance behavior for a chosen $B_1+$ value. The needed excitation bandwidth is approximately +-50 kHz (depicted as the shaded gray region in the bottom plots.) *(a,b)* Excitation and refocusing pattern for hard pulses (80 μs and 160 μs for excitation and refocusing respectively). The resulting profiles cover a narrow bandwidth (1-4 KHz) and are very sensitive to the B1 amplitude. *(c,d)* Excitation and refocusing pattern for frequency-swept WURST pulses (3.2 ms and 1.6 ms long for excitation and refocusing respectively, both with a 100 kHz sweep). These cover wider bandwidths and the refocusing pulse has a reduced sensitivity to the $B_1$ amplitude.

We use 3D RARE sequences, which support standard $T_2$, IR-prepped $T_2$, $T_1$, proton density (PD), and diffusion contrasts. Since no slice-selective gradient is employed, the system acquires 3D encoded axial imaging where the y phase encode gradient is used for partitioning the 3D data into ~7mm thick image partitions. Figure 4 depicts the basic pulse sequence diagram, including the chirped RF pulses, the $G_y$ phase-encoding blips (varying down the spin-echo train), the $G_z$ phase-encoding blips (incrementing shot-to-shot for each spin-echo train), and the $G_x$ permanent magnet read-out gradient (always on). The resulting spin-echoes (signal acquisition line) show the previously described alternating "FID-echo" and "spectral-echo" behanvior[27].

For $T_2$-weighting, the y-dimension (partition) phase encoding is performed along the echo train with a k-space trajectory placing the center of k-space in the middle echo. The Z-dimension gradient phase encoding is incremented shot to shot. The proton density sequence uses a "center-out" k-space ordering down the echo-train. The $T_1$ sequence is similar but includes an Inversion Recovery (IR) prep pulse.

***In vivo experiments:*** Subjects were setup in a supine position on the detachable patient table for imaging. Before attaching the patient table to the scanner, the RF coil was positioned on the subject. A 1-minute, low-resolution image acquisition was used to confirm the proper positioning of the subject's head in the coil and magnet. All *in vivo* images were acquired with a matrix size of 256 x 97 x 23 and an approximate resolution = 2.2 x 1.3 x 6.8 $mm^3$. Subject 1 was imaged with the 3D RARE PD-weighted sequence (TR/TEeff = 2900ms / 14ms, 2 averages, acquisition time = 9:24 min), the inversion-prepped 3D RARE T1-weighted sequence (TI/TR/TEeff = 400ms / 1830ms / 14ms, 4 averages, acquisition time = 11:46 min), and the 3D RARE T2-weighted sequence (TR/TEeff =3000ms/167ms, 2 averages, acquisition time = 9:42 min). Subject 2 was imaged with the 3D RARE T2-weighted sequence (TR/TEeff =3000ms/167ms, 2 averages, acquisition time = 9:42 min). Subject 3 was imaged with the 3D RARE T2-weighted sequence (TR/TEeff =3000ms/167ms, 4 averages, acquisition time = 19:24 min). The study was approved by the institutional review board of the Partners Healthcare, and written informed consent was obtained before scanning.

***Image reconstruction method***: The images are reconstructed from the data using a generalized encoding matrix model to describe the expected signal based on the measured field maps and an iterative linear solver to determine the image[29,32,33,30,22]. This provides a more accurate relation between the encoded signal and the object than the Fourier model (which assumes linear encoding fields) and, in principle, alleviates image distortion from the imperfect encoding fields.

A full 3D reconstruction can be done using all three encoding field-maps, but to reduce the matrix size the data is partitioned in the y direction using the FFT. The generalized reconstruction method is then used to reconstruct each 2D axial image. Specifically, the encoding matrix represents the phase at each time domain sample point in an echo, imparted by the $G_x$ readout encoding field and the $G_z$ phase encoding blips. Without relaxation effects, the assumed signal equation for the readout time point, t, and the n[th] $G_z$ phase encode is modeled as:

$$s_n(t) = \sum_r e^{-i2\pi\gamma(G_x(r)t + I(n)G_z(r)\tau)} m(r)$$

where *r* is the 2D position, *γ* is the gyromagnetic ratio in Hz/Tesla, $G_x$ is the non-linear built-in readout gradient field map (in units of Tesla), *I(n)* is the $G_z$ scaling factor for nth phase encode blip, $G_z$ is the 2D phase encoding gradient field map, *τ* is the length of the phase encode blip, and *m* is the image. A coil sensitivity weighting is not included because we assume a uniform receive sensitivity from the volume coil.

This equation can be simplified as a matrix-vector product, where the matrix contains the known field quantities ($G_x$ and $G_z$) and the vector is the list of image pixels to be estimated. Therefore, image reconstruction is a linear inverse problem that we solve using conjugate gradient (CG). The system matrix is very large in our case; therefore, we do not store in memory and instead compute its rows online, which is very fast. The $G_x$ and $G_z$ field maps contain a few million elements each, and therefore fit easily in the global shared memory of modern graphical processing unit (GPU) such as the Tesla K20c (5 GB) or the more recent Tesla P100 (16 GB). Implementation of the matrix-vector product (Ax and $A^H$x, where [H] is the complex-transpose operation) takes a couple of seconds in the GPU compared to ~1 hour on a single CPU. To minimize the total computation time, we also employ a preconditioner which is the diagonal matrix comprised of the square of the diagonal entries of the system matrix correlation matrix (C=$A^H$A). This is a good preconditioner for this problem since it is 1) ultra-rapid to compute, 2) is trivial to invert and 3) reduces the condition number of the problem from ~133 to ~2. As a result, iterative reconstruction of a 220x180 matrix-size image (FOV 22 x 18 cm) requires 5-10 iterations at the 0.1% convergence level which represents a total time of <20 seconds.

We apply an intensity correction to the images to alleviate shading caused by B1 inhomogeneity. This is done by masking each 2D image and dividing it by a low-pass-filtered version of itself. Image SNR calculations were performed on the FFT reconstructed version of the image, to reduce the confounding effects of noise amplification in iterative reconstruction. The calculation was performed in a central partition magnitude image. SNR was calculated as the mean of a high intensity ROI divided by the mean of background ROI. A factor of sqrt(pi/2) was applied to account for the Rician distribution of the magnitude image noise.

**Code availability.** The MATLAB code used to reconstruct the portable MRI data is available at https://github.com/czcooley/portable-MRI.

**Data availability.** The data that support the findings of this study are available within the paper. All datasets generated for this study are available from the corresponding author upon reasonable request.

**Acknowledgements**: We thank Thomas Witzel for valuable advice over the course of the system development and specific assistance with consoles. We thank Melissa Haskell for her contribution to the magnet design algorithm. We thank Matthias David for his assistance with gradient non-linearity analysis. We thank Simon Sigalovsky for the construction of mechanical components. We thank John Conklin for insightful discussions on clinical applications. We thank Neha Koonjoo her help with the helmet coil design. Research reported in this publication was supported by the National Institute of Biomedical Imaging and Bioengineering (NIBIB) of the National Institutes of Health under award number R01EB018976.

**Author contributions**: CZC, PCM, JPS, SAS, CRS, CFV, MS, MSR and LLW contributed to or advised on the system design, implementation, and validation experiments. CZC, JPS, SC, and BG contributed to the development of the image reconstruction method. MHL provided guidance for clinical application and subsequent design choices. CZC wrote the manuscript and all authors contributed to reviewing and editing.

*Submitted to Nature Biomedical Engineering*